\documentclass[10pt,aps,superscriptaddress,prl,a4paper,reprint, floatfix]{revtex4-1}

\usepackage[british]{babel}
\usepackage[utf8]{inputenc}

\usepackage{amssymb, amsmath, newtxtext}

\usepackage{graphicx}

\usepackage{xcolor}

\usepackage[per-mode=symbol]{siunitx}

\usepackage{hyperref}

\renewcommand{\vec}[1]{\mathbf{#1}}

\begin{document}
\title{Collective Modes of a Photon Bose-Einstein Condensate\\with Thermo-Optic Interaction}
\author{Enrico Stein}
\email{estein@rhrk.uni-kl.de}
\affiliation{Department of Physics and Research Center OPTIMAS, Technische Universität Kaiserslautern, Erwin-Schrödinger Straße 46, 67663 Kaiserslautern, Germany}

\author{Frank Vewinger}
\email{vewinger@iap.uni-bonn.de}
\affiliation{Institut für Angewandte Physik, Universität Bonn, Wegelerstraße 8, 53115 Bonn, Germany}

\author{Axel Pelster}
\email{axel.pelster@physik.uni-kl.de}
\affiliation{Department of Physics and Research Center OPTIMAS, Technische Universität Kaiserslautern, Erwin-Schrödinger Straße 46, 67663 Kaiserslautern, Germany}
\begin{abstract}
Although for photon Bose-Einstein condensates the main mechanism of the observed photon-photon interaction has already been identified to be of thermo-optic nature, its influence on the condensate dynamics is still unknown. Here a mean-field description of this effect is derived, which consists of an open-dissipative Schrödinger equation for the condensate wave function coupled to a diffusion equation for the temperature of the dye solution. With this system at hand, the lowest-lying collective modes of a harmonically trapped photon Bose-Einstein condensate are calculated analytically via a linear stability analysis. As a result, the collective frequencies and, thus, the strength of the effective photon-photon interaction turn out to strongly depend on the thermal diffusion in the cavity mirrors. In particular, a breakdown of the Kohn theorem is predicted, i.e.~the frequency of the centre-of-mass oscillation is reduced due to the thermo-optic photon-photon interaction.
\end{abstract}
\date{\today}
\maketitle

\emph{Introduction} -- In recent years many theoretical and experimental results have contributed to a basic understanding of quantum fluids of light \cite{RevModPhys.85.299}, where many photons propagate in nonlinear optical systems. The corresponding collective features are due to effective photon-photon interactions, which are induced by the nonlinear matter. The hydrodynamic behaviour of light in a cavity, first noted by Lugiato and Lefever in 1987 \cite{PhysRevLett.58.2209}, was theoretically brought forward in Ref.~\cite{PhysRevA.48.1573} by deriving a Ginzburg-Landau equation for laser light inside a cavity. These theoretical works were complemented by the experimental proof of superfluidity of light via the pioneering observation of stable quantised vortices by Swartzlander and Law in 1992 \cite{PhysRevLett.69.2503}. With this the natural question arose, whether light could also undergo the equilibrium phase transition of Bose-Einstein condensation. This intriguing question was partly answered in 2002 when the first exciton-polariton condensate was realised \cite{Science.298.5591}. However, such condensates have turned out to be not of a Bose-Einstein type, as their life time is shorter than the intrinsic equilibration time. In contrast to that an equilibrium Bose-Einstein condensate (BEC) of pure light was achieved in Bonn in 2010 \cite{Klaers2010a}. Although this is still a driven-dissipative system as the exciton-polariton condensates, the favourable timescale ratio allows for the observation of equilibrium effects \cite{Schmitt2015}.\\
The experimental setup to create a BEC of photons consists of a microcavity filled with a dye solution. There the cavity provides a well-defined ground state for the effective two-dimensional photon gas, as can be seen from the paraxial approximation, and the dye leads to a thermalisation of the photon gas via absorption and emission processes of the photons \cite{Klaers2010}. As the corresponding absorption and emission rates are related via a Boltzmann factor according to the Kennard-Stepanov relation \cite{Kennard1918,Kennard1926,Stepanov1957,Kazachenko1957}, the photon gas inherits the thermalisation from the dye molecules. Provided that the pumping power is large enough in order to compensate unavoidable cavity losses and thermalisation proceeds faster than the losses, the photon gas can undergo an equilibrium Bose-Einstein phase transition \cite{Klaers2011}.\\
Furthermore, the absorption and emission processes lead to effective photon-photon interaction mechanisms, see Fig.~\ref{fig1}. One is the Kerr effect, where a nonlinear susceptibility causes the refractive index of the dye solution to be proportional to the intensity of the electric field in the cavity \cite{Boyd}. A microscopic theory of the Kerr interaction in a photon BEC is based on a Lindblad master equation \cite{Kirton2013,Radonjic2018}. A second interaction effect is due to the heating of the dye solution, as the quantum efficiency of the dye is below 100 \%. This leads to a shift of the refractive index of the solvent \cite{Boyd} and correspondingly to a thermo-optic photon-photon interaction. As the latter is mediated by the temperature diffusing through the dye solution, it is nonlocal in space and retarded in time.
\begin{figure}
	\includegraphics[width=.75\linewidth]{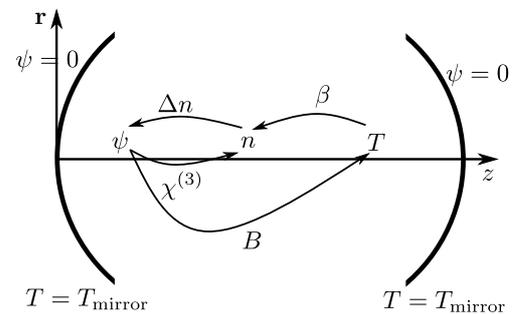}
\caption{\label{fig1}Scheme of effective photon-photon interaction mechanisms. The electric field, here represented by the condensate wave function $\psi$, couples in two ways to the refractive index $n$. Once via the Kerr effect, which is due to the non vanishing third-order susceptibility $\chi^{(3)}$, and, secondly, also an increase of the dye-solution temperature $T$ due to the heating coefficient $B$ is present, which then yields via the thermo-optic coefficient $\beta$ a change $\Delta n$ of the refractive index.}
\end{figure}
So far the strength of the effective photon-photon interaction has been experimentally determined by measuring the increase of the condensate width with the photon number \cite{Klaers2010a, Klaers2011}. From this it is concluded, that the main contribution of the interaction is due to the thermo-optic effect. The strength of the interaction, which can be defined to be dimensionless in two spatial dimensions, is measured to have values up to $\tilde{g}=\num{7(3)e-4}$ \cite{Klaers2010a, Klaers2011}, where the precise value depends on the detailed experimental configuration. An interaction strength of the same order was observed in another experiment in London in 2016 \cite{Marelic2016}.\\
From atomic BECs it is known that observing the condensate dynamics represents a valuable diagnostic tool to measure system properties in general and two-particle interaction strengths in particular. For instance, observing collective frequencies of trapped condensates is a precise way to measure the strength of the contact interaction up to an astonishing precision of \mbox{1 \textperthousand} \cite{PhysRevLett.81.500,Pollack2010}. Therefore, observing the collective frequencies of a photon BEC is expected to yield additional profound information about the nature and the strength of the effective photon-photon interaction.
This motivates to analyse in the following the lowest-lying collective modes of a photon BEC under the influence of the thermo-optic interaction in view of future experiments.

\emph{Model} -- A minimal mean-field description of the thermo-optic interaction consists of two equations \cite{PhysRevLett.113.135301,Alaeian2017, Dung2017}. One is a nonlinear Schrödinger equation that accounts for the evolution of the electric field inside the cavity, which is assumed to be linearly polarized, and the second equation describes the diffusion of the temperature, produced by the non-perfect absorption processes of photons. As the experiment takes place inside a microcavity, the electric field can be treated in paraxial approximation \cite{PhysRevA.11.1365, Boyd} which allows to map the three-dimensional massless photon gas in a spherical mirror geometry to a two-dimensional gas of bosonic particles. These particles possess a mass $m=\hbar\omega_\text{cutoff}(c/n_0)^2$, where the cavity-cutoff frequency is denoted by $\omega_\text{cutoff}$ and the light velocity in the dye solution is $c/n_0$. Furthermore, they are trapped in a harmonic potential with frequency $\Omega=c\sqrt{2/(L_0R)}/n_0$, that is determined by the cavity length $L_0$ and the radius of curvature $R$ of the mirror \cite{Klaers2010a,Klaers2011,Nyman2018}. Thus, the evolution of the condensate wave function $\psi(\vec{r},t)$, i.e.~the electric field normalised to the photon number, is described by an open-dissipative Schrödinger equation of the form \cite{Alaeian2017,PhysRevA.90.043853,Nyman2018,Stein2018}
\begin{align}\label{eq1}
	i\hbar\partial_t \psi(\vec{r},t)=&\left\lbrace-\frac{\hbar^2\nabla^2}{2m}+\frac{m\Omega^2}{2}\vec{r}^2+g_T\Delta T(\vec{r},t)\right.\\
	&\left.+\frac{i\hbar}{2}\left[p-\Gamma+\frac{p+\Gamma}{n_0}\beta \Delta T(\vec{r},t)\right]\right\rbrace\psi(\vec{r},t).\nonumber
\end{align}
Here, the remaining transversal degrees of freedom are denoted by $\vec{r}=(r_1,r_2)^T$. The thermo-optic effect is described by the nonlinearity in Eq.~(\ref{eq2}) involving the temperature difference $\Delta T(\vec{r},t)$ between the actual intra cavity temperature and the room temperature. Here the coupling coefficient $g_T=-\beta mc^2/n_0$ quantifies the energy shift due to the heating. As discussed above, it is justified to neglect the much smaller Kerr interaction. Due to the unavoidable cavity losses, the photon BEC is intrinsically an open system. Following Ref.~\cite{PhysRevLett.99.140402}, an incoherent pump scheme is modelled by the imaginary part in Eq.~(\ref{eq2}). The pump is described by the coefficient $p$ and the losses by the decay rate $\Gamma$. Note that the emission and absorption processes, which do not lead to a loss of photons, i.e.~the coherent ones that are proportional to the quantum efficiency $\eta$, are not considered here. Thus, the loss rate $\Gamma$ is only proportional to $1-\eta$ and gives rise to a heating of the dye solution as discussed below.\\
On the other hand the temperature difference $\Delta T(\vec{r},t)$ follows a diffusion equation. Reducing it from three to two spatial dimensions by the procedure described in Ref.~\cite{Suppl} yields
\begin{equation}\label{eq2}
	\partial_t \Delta T(\vec{r},t)=\left(D_0\nabla^2-\frac{1}{\tau}\right)\Delta T(\vec{r},t)+B|\psi(\vec{r},t)|^2,
\end{equation}
where the temperature diffusion constant is denoted by $D_0=\lambda_w/(c_p\rho)$. It depends on the thermal conductivity $\lambda_w$, the specific heat $c_p$ and the density $\rho$ of the solvent \cite{Landau2009}. The heating coefficient of the dye solution is given by $B=mc^2\Gamma/(L_0n_0c_p\rho)$ \cite{Dung2017}. The temperature relaxation is governed by the time scale  $\tau$, which depends in general on length scales of both the cavity and the mirrors \cite{Suppl}. Note that in Ref.~\cite{PhysRevLett.113.135301} only the limiting case $\tau\rightarrow\infty$ is treated.\\
A similar mean-field model was already established within the realm of exciton-polariton condensates \cite{PhysRevLett.99.140402}, where the exciton bath plays a role comparable to the temperature for the photon BEC. However, the time scales of the two systems are inverted. In an exciton-polariton condensate the relaxation of the exciton reservoir is fast compared to the dynamics of the condensate, allowing its adiabatic elimination. In contrast to that, the photon BEC dynamics, which is determined by the trap frequency $\Omega$, occurs on a much faster time scale than the dynamics of the temperature, whose time scale is set by the large relaxation time $\tau$. Thus, the resulting thermo-optic photon-photon interaction yields such a significant temporal retardation that no influence on any condensate dynamics is expected.\\
Nevertheless, in the following it is shown that the collective frequencies of the photon BEC turn out to be modified by the thermo-optic photon-photon interaction. The reason is that, in the steady state, the refractive index near the trap centre is modified, so the collective modes exploring its neighbourhood get changed. Or, put differently, the temperature profile can be considered as the motional history of the condensate and, thus, the condensate effectively scatters with its own history.

\emph{Method} -- 
The usual variational approach for dealing with collective excitations in ultracold quantum gases is based on Hamilton's principle \cite{PhysRevLett.77.5320, PhysRevA.56.1424, PhysRevA.57.1191}. As the photon BEC is intrinsically an open system, however, no such principle exists, as the energy is not a conserved quantity. This problem can be circumvented by considering the equations of motion of the cumulants \cite{PhysRevA.56.2978,PhysRevLett.120.063605}, i.e. calculating the evolution equations for the centre-of-mass and for the widths. Due to the openness of the system also the photon number $N(t)=\int d^2r~|\psi(\vec{r},t)|^2$ represents an additional variational parameter. The ansatz for the condensate wave-function is chosen to be of Gaussian shape, as this is a solution to the non interacting and closed version of the Schrödinger equation (\ref{eq2}):
\begin{align}\label{eq3}
	\psi(\vec{r},t)=&\sqrt{\frac{\sigma N(t)}{\pi q_1(t)q_2(t)}}\exp\Bigg\lbrace\sum_{j=1,2}-\left[\frac{1}{2q_j(t)^2}+iA_j(t)\right]\nonumber\\&\times\left[r_j-x_{0j}(t)\right]^2+ir_jC_j(t) \Bigg\rbrace,
\end{align}
where the centre-of-mass coordinates are denoted by $x_{0j}(t)$ with the phases $C_j(t)$, whereas $q_j(t)$ describe the condensate widths, and the $A_j(t)$ stand for the corresponding phases. Since all current photon BEC experiments are working with a particular pump sequence \cite{Klaers2010a, Klaers2011}, the duty cycle $\sigma$ models a continuous pump in the ansatz. Thus, the number of photons present on average in the cavity is given by $\sigma N$. This averaging coarse-grains the dynamics, and for the steady-state values considered later the ansatz turns out to be reasonable. As the temperature difference is pumped by the photons, it is justified to assume also a Gaussian shape for its distribution, which by itself solves the homogeneous part of the diffusion equation (\ref{eq1}):
\begin{align}\label{eq4}
	\Delta T(\vec{r},t)=\frac{\Delta T_0(t)}{\pi s_1(t)s_2(t)}\exp\left\lbrace\sum_{i=1,2}-\frac{\left[r_i-y_{0i}(t)\right]^2}{s_i(t)^2} \right\rbrace.
\end{align}
Here $\Delta T_0(t)$ denotes the amplitude of the temperature difference, $y_{0i}(t)$ describes the centres of the distribution and $s_i(t)$ their widths. The aim is now to calculate the equations of motion for the amplitudes, the centre-of-masses and the widths, as well as their phases, following the procedure of Ref.~\cite{PhysRevLett.120.063605}. The resulting system of coupled ordinary differential equations is shown in detail in Ref.~\cite{Suppl}.

\emph{Steady State} -- 
\begin{figure}
	\includegraphics[width=.9\linewidth]{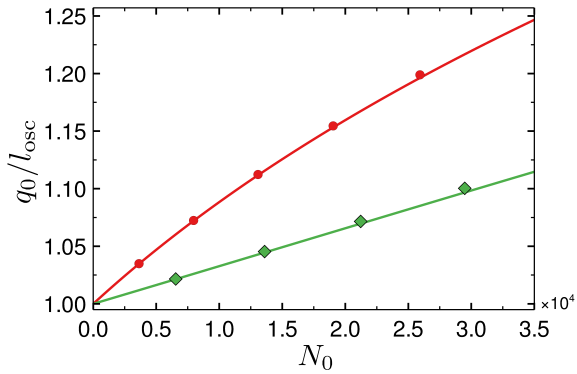}
	\caption{\label{fig2}Steady state (\ref{eq5}) of the system (\ref{eq1}), (\ref{eq2}) with the ansatz (\ref{eq3}), (\ref{eq4}) for two different values of the diffusion constant $D$. The red line (circles) corresponds to the experimental value $D_\text{exp}=\SI{9.16e-8}{\square\meter\per\second}$ and the green one (diamonds) to the larger value $D=10^2D_\text{exp}$. The markers indicate the full numerical solution of the system (\ref{eq1}) and (\ref{eq2}), which is based on Refs. \cite{Muruganandam2009,Vudragovic2012,Young-S.2016}.}
\end{figure}
From the equations of motions of the cumulants, see \cite{Suppl}, it follows that the temperature evolution is determined by the condensate, so the steady state is completely described by the latter. Furthermore, due to the trap isotropy the condensate shape is also isotropic. Accordingly, the dependence of the equilibrium condensate width $q_0$ on the equilibrium photon number $N_0$ is described by
\begin{align}\label{eq5}
	0=\frac{1}{q_0^4}-\frac{1}{l_\text{osc}^4}+\frac{2\tilde{g}N_0}{\pi(2q_0^2+D_0\tau)^2},
\end{align}
where $l_\text{osc}=\sqrt{\hbar/(m\Omega)}$ denotes the oscillator length. Moreover, the dimensionless interaction strength turns out to be $\tilde{g}=m\sigma g_T\tau B/\hbar^2$, showing that the interaction strength is determined by the properties of the used dye solution, the geometry of the microcavity, as well as by the pump scheme. Note, that this result is comparable to that in Ref.~\cite{Dung2017}. In addition, the effective photon-photon interaction can directly be controlled via the relaxation time $\tau$ by changing the geometry of the cavity mirrors. It turns out that in the experimental situation, where the longitudinal and the transversal extension of the mirrors are of the order $L_1\sim\SI{1}{\centi\meter}$ and $L_\perp\sim\SI{1}{\milli\meter}$, the temperature relaxation inside the cavity is governed by the transversal temperature diffusion in the mirrors, see Ref.~\cite{Suppl}, yielding 
\begin{equation}
    \tau = \frac{L_\perp^2}{2\pi^2 D_1}.
\end{equation}
In case of a large diffusion constant of the solvent, i.e. $D_0\tau\gg q_0^2$, the condensate width $q_0$ approaches the noninteracting value $l_\text{osc}$. This results from a suppression of the thermo-optic interaction as all the temperature excitations are quickly transported through the dye solution. In the opposite case the behaviour $q_0=l_\text{osc}\left[1+N_0\tilde{g}/(2\pi)\right]^{1/4}$ is reproduced, which is well known from 2D atomic BECs \cite{Ghosh2002}.\\
The numerical solution of Eq.~(\ref{eq5}) is shown in Fig.~\ref{fig2}, where the values of the Bonn experiment with the solvent ethylene glycol are used \cite{Klaers2011,nyman_robert_andrew_2017_569817, ethyleneglycol}. The dye solution is characterised by $n_0=1.46,~\lambda_w=\SI{0.26}{\watt\per\meter\per\kilogram},~c_p=\SI{144.5}{\joule\per\mol\per\kelvin},~\rho=\SI{1110}{\kilogram\per\cubic\meter},~\beta=\SI{-4.68e-4}{\per\kelvin}$ and $\Gamma=\SI{1}{\per\second}$. The cavity geometry is given by $L_0=\SI{1.5e-6}{\meter}~\text{and}~R=\SI{1}{\meter}$. The pump scheme yields a duty cycle of $\sigma=\num{1/16000}$. The effective mirror diffusion constant $D_1$ is of the order of \mbox{$10^{-6}$ \si{\square\meter\per\second}}, thus yielding the relaxation time $\tau\sim\SI{0.1}{\second}$. With this the value of the dimensionless interaction constant is calculated within the mean-field model to be $\tilde{g}\sim 10^{-4}$, which is in remarkable agreement with the experimental value \cite{Klaers2010a, Klaers2011, Marelic2016}. 

\emph{Linearised Dynamics} --
Linearising the equations of motion with respect to small elongations out of the equilibrium yields a decoupling of the centre-of-mass from the width dynamics. Therefore, the dipole mode, which is a pure centre-of-mass motion, can be discussed separately from the breathing and the quadrupole mode, which are in- or out-of-phase width oscillations, respectively. At first the investigation aims for the dipole mode, which can be described by the vector $\vec{v}_i=(\delta x_{0i},\delta\dot{x}_{0i},\delta y_{0i})^T$, where $\delta$ denotes small perturbations of the steady state. The linearised equation of motion is given by $\dot{\vec{v}}_i=S_\text{dipole}\vec{v}_i$.
\begin{figure}
	\includegraphics[width=.9\linewidth]{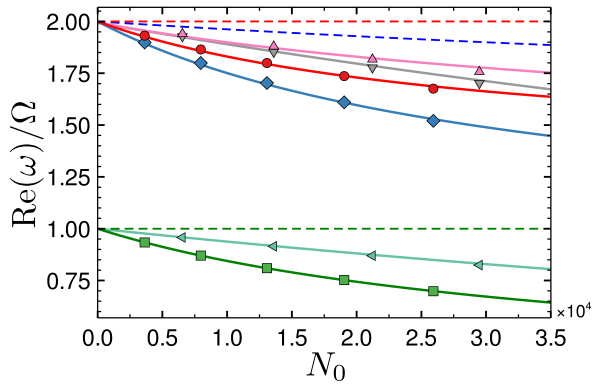}
	\caption{\label{Fig3}Oscillation frequencies obtained from real parts of the eigenvalues of $S_\text{dipole}$ describing the dipole mode oscillation (green, squares) and of $S_\text{widths}$ describing the quadrupole (blue, diamonds) and the breathing mode (red, circles) for the experimental value of the diffusion constant $D_{0\text{exp}}=\SI{9.16e-8}{\square\metre\per\second}$. For the larger diffusion constant $D_0=10^2D_\text{exp}$ the dipole mode frequency is visualised by the light blue (left triangles), the one of the quadrupole mode by the grey line (lower triangles) and the breathing mode by the pink (upper triangles) line. The dashed ones result from a variational evaluation of a simple Gross-Pitaevskii equation with a harmonic trapping potential. The markers visualise the results of a numerical evaluation of system (\ref{eq1}) and (\ref{eq2}), which is based on Refs. \cite{Muruganandam2009,Vudragovic2012,Young-S.2016}.}
\end{figure}
The eigenvalues of $S_\text{dipole}$ are complex due to the pumping and the coupling to the diffusion equation. The real parts describe the oscillation frequency of the dipole mode, whereas the imaginary parts represent the corresponding damping rates, which are discussed in Ref.~\cite{Suppl}. The frequencies are shown in Fig.~\ref{Fig3}, where the dashed lines are the results of a variational solution of the plain Gross-Pitaevskii equation \cite{PhysRevLett.77.5320}. In atomic BECs the dipole mode frequency equals the trap frequency, which corresponds to a centre-of-mass motion in a harmonic potential, according to the Kohn theorem \cite{PhysRevA.57.1191}. In the present case of a temporal nonlocal interaction, however, a shift to smaller frequencies is observed, which shows that the aforementioned scattering of the condensate with its own history leads, indeed, to a slowing down of the condensate motion and, thus, to the breakdown of the Kohn theorem. Furthermore, a stronger temperature diffusion in the solvent leads to a smaller frequency shift. This shows again that the diffusion suppresses the thermo-optic interaction.\\
The equations of motion describing the breathing and the quadrupole mode are coupled equations of the temperature and the condensate widths in both directions as well as the photon number and the temperature amplitude, which are summarised in the vector $\vec{w}=(\delta T,\delta N, \delta q_1, \delta \dot{q}_1, \delta r_1, \delta q_2, \delta \dot{q}_2, \delta r_2)^T$. The evolution of these quantities is described by $\dot{\vec{w}}=S_\text{widths}\vec{w}$, where the real part of the eigenvalues of $S_\text{widths}$ is shown in Fig.~\ref{Fig3}. Again, the frequencies are shifted to smaller values compared to a contact interaction. Even the breathing mode frequency, which turns out to be always twice the trap frequency for a contact interaction irrespective of the particle number and the strength of the contact interaction \cite{Ghosh2002}, gets shifted to smaller values. As before, we find that a larger diffusion strength yields a smaller frequency shift.\\
Analysing the corresponding damping rates leads to non-physical values due to
the missing matter degree of freedom in the current mean-field model \cite{Suppl}.
However, the damping rates are expected to be of the order of the reabsorption time, i.e.~of the order of \SIrange{10}{100}{\pico\second} \cite{Schmitt2015}. Thus, in view of a trap frequency of $\Omega=2\pi\times\SI{37}{\giga\hertz}$, a few oscillations should be experimentally observable.

\emph{Summary and Experimental Perspective} -- In this paper the influence of a thermal shift of the refractive index on the photon BEC dynamics is worked out. This shift yields an effective photon-photon interaction which is nonlocal in space and retarded in time. Due to the geometry dependence of the temperature diffusion, the strength of the effective photon-photon interaction can be controlled by the shape of the mirrors. Moreover, due to the retardation in time, the Kohn theorem does not hold in the present case and the dipole mode frequency is shifted to frequencies smaller than the trap frequency. Additionally, the temperature diffusion of the solvent has a large influence on the effective photon-photon interaction. The same happens to the breathing and the quadrupole mode. However, the damping rates of the collective modes are supposed to be larger once a more detailed theory takes the absorption and emission behaviour of the dye into account.\\
The above predicted features can, in principle, be measured in two ways. The first one relies on a direct observation of the collective modes. The dipole mode, e.g., can be excited by using two lasers, where the first one pumps the cavity homogeneously and the second one creates a BEC via an off-centre pulse. The excited mode can then be observed by measuring spatially the light leaking out of the cavity. As the dipole mode oscillation frequencies are expected to be of the order of the trap frequency, a streak camera is necessary in order to resolve these extreme time scales. The experiment performed in Ref.~\cite{Schmitt2015} can be seen as a proof of principle in this respect. The second method is an indirect measurement via the eigenfrequencies of the cavity \cite{PrivNyman}. In case of small interaction, which is the case in the photon BEC, the dipole mode corresponds to the difference of the lowest two cavity eigenfrequencies. Due to the interaction and the corresponding condensate broadening the third lowest energy state is also partially populated. This allows to examine the breathing-mode by spectrally resolving the cavity emission. 

\emph{Acknowledgement} -- We thank Hadiseh Alaeian, Antun Bala\v{z}, Erik Busley, Wassilij Kopylov, Christian Kurtscheid, Robert Nyman, Milan Radonjić, Julian Schmitt, Dries van Oosten, Georg von Freymann and Martin Weitz for interesting and useful discussions. Furthermore, we  acknowledge support and funding by the Deutsche Forschungsgemeinschaft (DFG, German Research Foundation) - Project number 277625399 - TRR 185.

\newpage

\title{Collective Modes of a Photon Bose-Einstein Condensate\\with Thermo-Optic Interaction\\Supplemental Material}
\author{Enrico Stein}
\email{estein@rhrk.uni-kl.de}
\affiliation{Department of Physics and Resarch Center OPTIMAS, Technische Universität Kaiserslautern, Erwin-Schrödinger Straße 46, 67663 Kaiserslautern, Germany}

\author{Frank Vewinger}
\email{vewinger@iap.uni-bonn.de}
\affiliation{Institut für Angewandte Physik, Universität Bonn, Wegelerstraße 8, 53115 Bonn, Germany}

\author{Axel Pelster}
\email{axel.pelster@physik.uni-kl.de}
\affiliation{Department of Physics and Resarch Center OPTIMAS, Technische Universität Kaiserslautern, Erwin-Schrödinger Straße 46, 67663 Kaiserslautern, Germany}

\date{\today}
\maketitle

\onecolumngrid
\setcounter{figure}{0}
\setcounter{equation}{0}
\renewcommand{\theequation}{S\arabic{equation}}
\renewcommand{\thefigure}{S\arabic{figure}}
\appendix

\begin{center}
	\large\bf
	Temperature Diffusion in Mirrors
\end{center}

\begin{figure}[b]
	\includegraphics{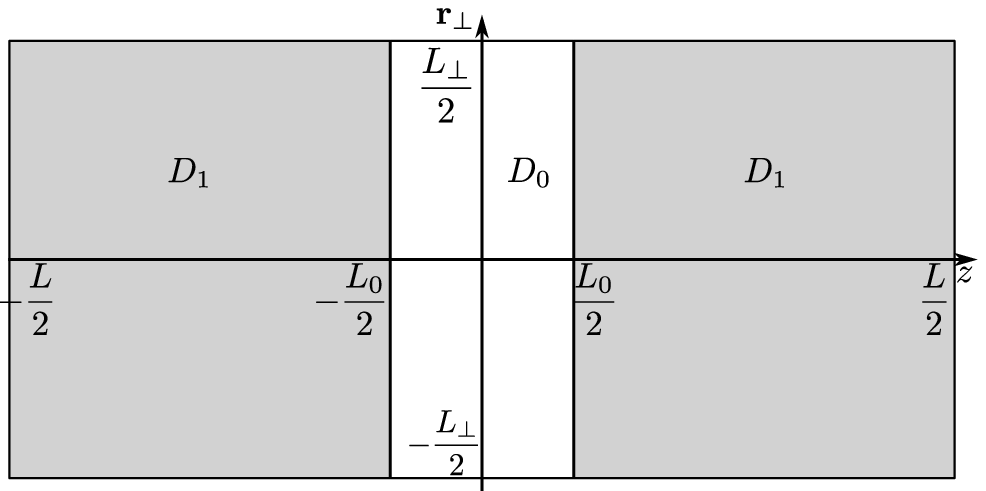}
	\caption{\label{figA1}Simplified geometry of the cavity setup with flat mirrors. Within the region $-L_0/2\leq z\leq L_0/2$ the diffusion constant takes the value $D_0$, whereas in the outer region $L_0/2\leq |z|\leq L/2$ the diffusion constant is given by $D_1$.}
\end{figure}

The purpose of this supplemental material is to derive the diffusion behaviour of the temperature within the cavity setup depicted in Fig.~\ref{figA1}. The geometry is simplified by assuming planar mirrors in comparison to the original geometry, which consists of spherically curved mirrors. 
\begin{center}
	\bf(a) Formulation of Boundary Value Problem
\end{center}
The complete cavity ranges from $z=-L/2$ to $z=L/2$, whereas the mirrors extend from the boundaries up to $z=-L_0/2$ up to $z=L_0/2$, respectively. The temperature diffusion constants of the dye solution filling the space between the mirrors and of the mirrors themselves are denoted by $D_0$ and $D_1$, respectively. Thus, the corresponding diffusion equation for the temperature difference $\Delta T$ between the actual temperature of the experimental setup and the room temperature reads \cite{Murray2002a}
\begin{align}\label{eq10}
	\partial_t \Delta T=\nabla\cdot\left[D(z)\nabla\Delta T~\right]+S,
\end{align}
with the function $D(z)=D_0+(D_1-D_0)\theta(|z|-L_0/2)$ containing the respective diffusion constants and $S$ being a source term with support only inside the cavity. 
Due to the symmetry of the considered geometry in Fig.~\ref{figA1}, it is sufficient to consider only the non-negative $z$ half space together with the von Neumann boundary condition
\begin{equation}\label{neumann_bound}
	\left.\partial_z\Delta T\right|_{z=0}=0.
\end{equation}
On the other hand, the temperature difference obeys the Dirichlet condition 
\begin{equation}
	\Delta T(z=L/2)=0
\end{equation}
at the border of the mirror. Furthermore, at $z=L_0/2$, i.e.~the contact of the mirror and the dye solution, the temperature difference is continuous:
\begin{equation}\label{eq9a}
	\lim_{\epsilon\rightarrow0}\Delta T\left(\frac{L_0}{2}-\epsilon\right)=\lim_{\epsilon\rightarrow0}\Delta T\left(\frac{L_0}{2}+\epsilon\right).
\end{equation}
Integrating Eq.~\eqref{eq10} in the neighbourhood of the material transition yields a jump condition for the first derivative of the temperature difference
\begin{align}\label{eq9}
	D_0\lim_{\epsilon\rightarrow 0}\left.\partial_z\Delta T\right|_{z=L_0/2-\epsilon}=D_1\lim_{\epsilon\rightarrow 0}\left.\partial_z\Delta T\right|_{z=L_0/2+\epsilon}.
\end{align}

\begin{center}
	\bf(b) Dimensional Reduction of Boundary Value Problem
\end{center}
The aim is now to derive an effective equation for the transversal diffusion of the temperature difference within the cavity. Due to the piecewise defined diffusion function $D(z)$, the ansatz for the temperature difference is chosen to be
\begin{equation}\label{ansatz_inner}
	\Delta T_0 = \Delta T_{\perp0}(\vec{r}_\perp,t)\Delta T_{\parallel0}(z)
\end{equation}	
inside the cavity and
\begin{equation}\label{ansatz_mirror}
	\Delta T_1 = e^{-t/\tau}\Delta T_{\perp1}(\vec{r}_\perp)\Delta T_{\parallel1}(z)
\end{equation}
for the mirror. By writing down these two ansatzes three assumptions have been made. First, the transversal component inside the cavity $\Delta T_{\perp0}$ varies on a time scale set by the photon condensate. This scale is much faster than the intrinsic scale of the diffusion process and, therefore, this component acts on its own time scale. The second assumption accounts for the steady state of the diffusion process in the mirror. Accordingly, the ansatz \eqref{ansatz_mirror} involves only an exponential time dependence with the relaxation time $\tau$. Lastly, only the lowest temperature difference mode is considered, as this mode decays least and has, therefore, the largest amplitude. With the two ansatzes \eqref{ansatz_inner}, \eqref{ansatz_mirror} Eq.~\eqref{eq10} reduces to one diffusion equation for each region, which are linked via the boundary conditions \eqref{neumann_bound}--\eqref{eq9}.

\begin{figure}
    \centering
    \includegraphics[width=.4\linewidth]{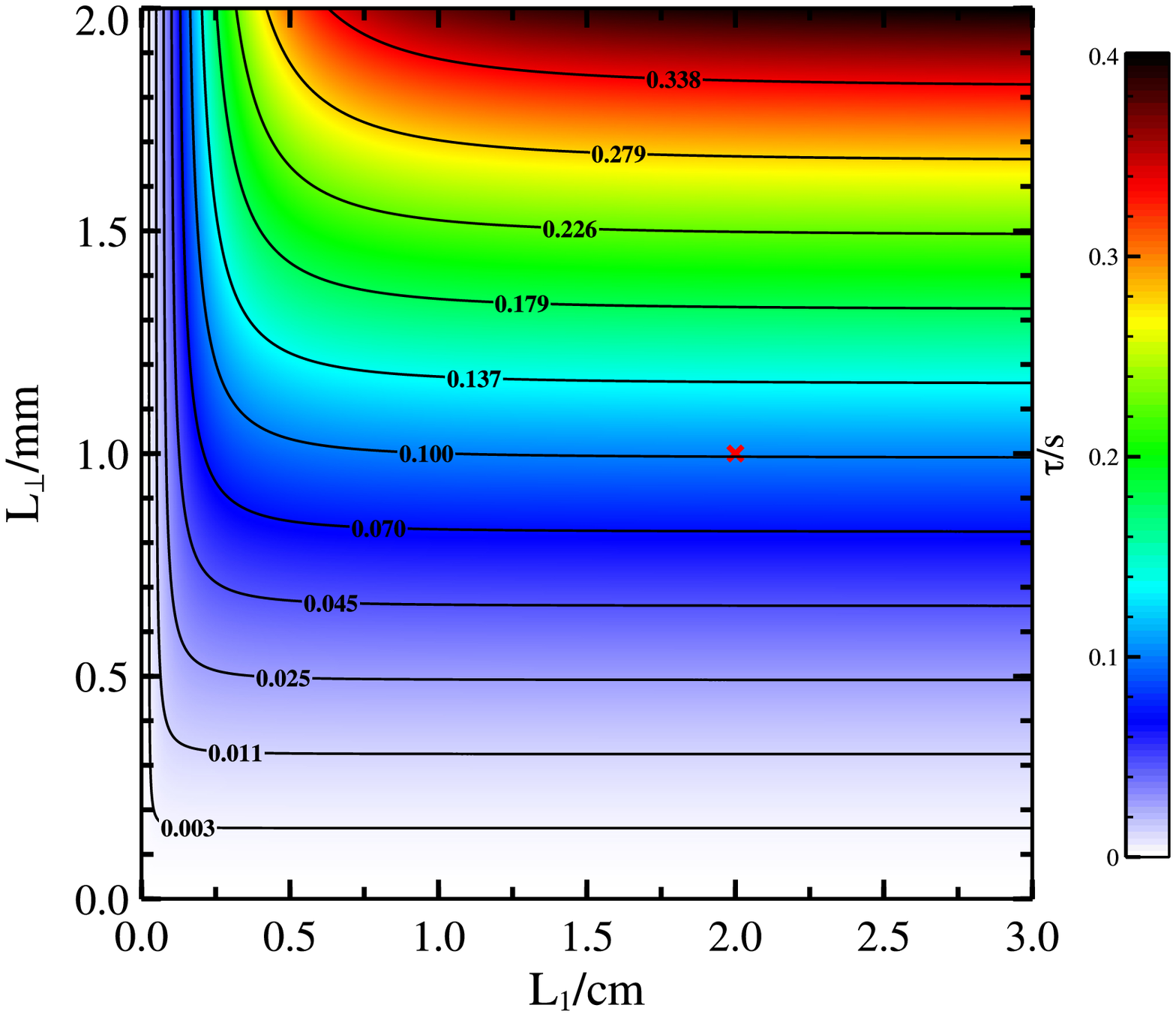}
    \caption{Plot of the relaxation time \eqref{tau_f} for the parameter region relevant in the Bonn experiment. The contour lines and the colour code provide the value of $\tau$. The red cross marks the current values of the Bonn experiment.}
    \label{figA4}
\end{figure}

Solving the diffusion boundary value problem leads to the following results.
The relaxation time $\tau$ in the mirrors consists of the decay time of the transversal diffusion process
\begin{equation}
	\tau_{\perp} = \frac{L_\perp^2}{4\pi^2D_1}
\end{equation}
and the longitudinal relaxation time $\tau_{\parallel}$ which is determined by the transcendental equation
\begin{equation}
	\sqrt{\frac{D_0}{\tau}}\tan\left(\frac{L_0/2}{\sqrt{\tau D_0}}\right) = \sqrt{\frac{D_1}{\tau_{\parallel}}}\cot\left(\frac{L_1/2}{\sqrt{\tau_{\parallel}D_1}}\right).
\end{equation}
Here, the mirror length is denoted by $L_1 = L-L_0$.
Furthermore, we assume that the relaxation time of the longitudinal diffusion inside the cavity occurs at the same time scale as the diffusion process inside the mirrors. In the limit of a microcavity, i.~e.~$L_0\ll L_1$, one finds
\begin{equation}
	\tau_{\parallel1}=\frac{L_1^2}{\pi^2D_1}.
\end{equation}
Therefore, the total mirror decay time $\tau$ is given by
\begin{equation}\label{tau_f}
	\frac{1}{\tau}=\frac{4\pi^2D_1}{L_\perp^2}+\frac{\pi^2D_1}{L_1^2}.
\end{equation}
This result is plotted in Fig.~\ref{figA4} for the parameters of the Bonn experiment, where the width of the mirrors is given by $L_\perp=\SI{2}{\milli\meter}$, whereas the length of the mirror $L_1$ is of the order of $\SI{1}{\centi\meter}$. Thus, in that cases \eqref{tau_f} simplifies to
\begin{equation}
	\tau = \frac{L_\perp^2}{2\pi^2 D_1}.
\end{equation}
As a consequence, the effective longitudinal relaxation time of the temperature difference inside the cavity is provided by the transversal temperature difference diffusion in the mirrors. Finally, the resulting two-dimensional diffusion equation for the transversal temperature difference within the cavity turns out to be
\begin{align}
    \partial_t \Delta T_{\perp0} = \left(D_0\nabla_\perp^2-\frac{1}{\tau}\right) \Delta T_{\perp0} + S_\perp,
\end{align}
which is Eq.~(2) in the paper. Note that the remaining source term takes the form
\begin{equation}
	S_\perp(\vec{r}_\perp,t) = \frac{2}{L_0}\int_0^{L_0/2}S(\vec{r}_\perp,z,t) \Delta T_{\parallel0}.
\end{equation}

\newpage
\vspace*{5mm}
\begin{center}
\large\bf
Detailed Equations of Motion
\end{center}
The mean-field equations (1), (2) are solved with the ansatz (3), (4) for condensate wave function and temperature difference by applying the cumulant approach \cite{PhysRevLett.120.063605}.
After eliminating the phases, the evolution of the remaining dynamical variables is described by
\begin{subequations}\label{eq_system}
	\begin{align}
		\partial_t{N}=&\left[p-\Gamma+\beta \Delta T_0\left(p+\Gamma\right)G_{T\psi}\right]N,\\
		\partial_t^2{x}_{0i}=&\partial_t(R_iq_i^2)+\left(\partial_t \ln(q_i)-I_iq_i^2\right)R_i-\Omega^2x_{0i}-2\Delta T_0G_{T\psi}\frac{g_T(y_{0i}-x_{0i})}{m(q_i^2+s_i^2)},\\
		\partial_t{q}_i=&q_i\partial_t(I_iq_i^2)-I_i^2q_i^5+2(\partial_t{q}_i)q_i^2I_i+\frac{\hbar^2}{m^2q_k^3}-\Omega^2q_k-\frac{4g_T}{m}G_{T\psi}\Delta T_0\left[\frac{(y_{0i}-x_{0i})^2q_i}{(s_i^2+q_i^2)^2}-\frac{q_i}{2(q_i^2+s_i^2)} \right],\\
		\partial_t{\Delta T}_0=&-\frac{\Delta T_0}{\tau}+\sigma BN,\\
		\partial_t{y}_{0i}=&\frac{\sigma BN}{\Delta T_0}(x_{0i}-y_{0i}),\\
		\partial_t^2{s^2_i}=&4D+\frac{\sigma BN}{\Delta T_0}(q_i^2-s_i^2)+2\frac{\sigma BN}{\Delta T_0}(x_{0i}-y_{0i})^2.
	\end{align}
\end{subequations}
Here, the overlap of the temperature-difference distribution and the condensate is described by 
\begin{equation}
	G_{T\psi}=\frac{\exp\left[-\sum_{i=1,2}\frac{(x_{0i}-y_{0i})^2}{s_i^2+q_i^2}\right]}{\left[\pi\prod_{i=1,2}\sqrt{\left(q_i^2+s_i^2\right)}\right]},
\end{equation}
and 
\begin{equation}
	I_i=\frac{2\beta T_0 G_{T\psi}}{n_0}\left[\frac{(y_{0i}-x_{0i})^2}{(s_i^2+q_i^2)^2}-\frac{1}{2(s_i^2+q_i^2)}\right]
\end{equation}
accounts for the pump and loss influence on the widths and 
\begin{equation}
	R_i=-p\frac{x_{0i}}{q_i^2+s_i^2}+\beta\Delta  T_0G_{T\psi}(p+\Gamma)\frac{y_{0i}-x_{0i}}{n_0(q_i^2+s_i^2)}
\end{equation}
are the corresponding ones for the centre-of-mass.

\vspace*{5mm}
\begin{center}
\large\bf
Damping Rates
\end{center}

\begin{figure}
	\includegraphics[width=.5\linewidth]{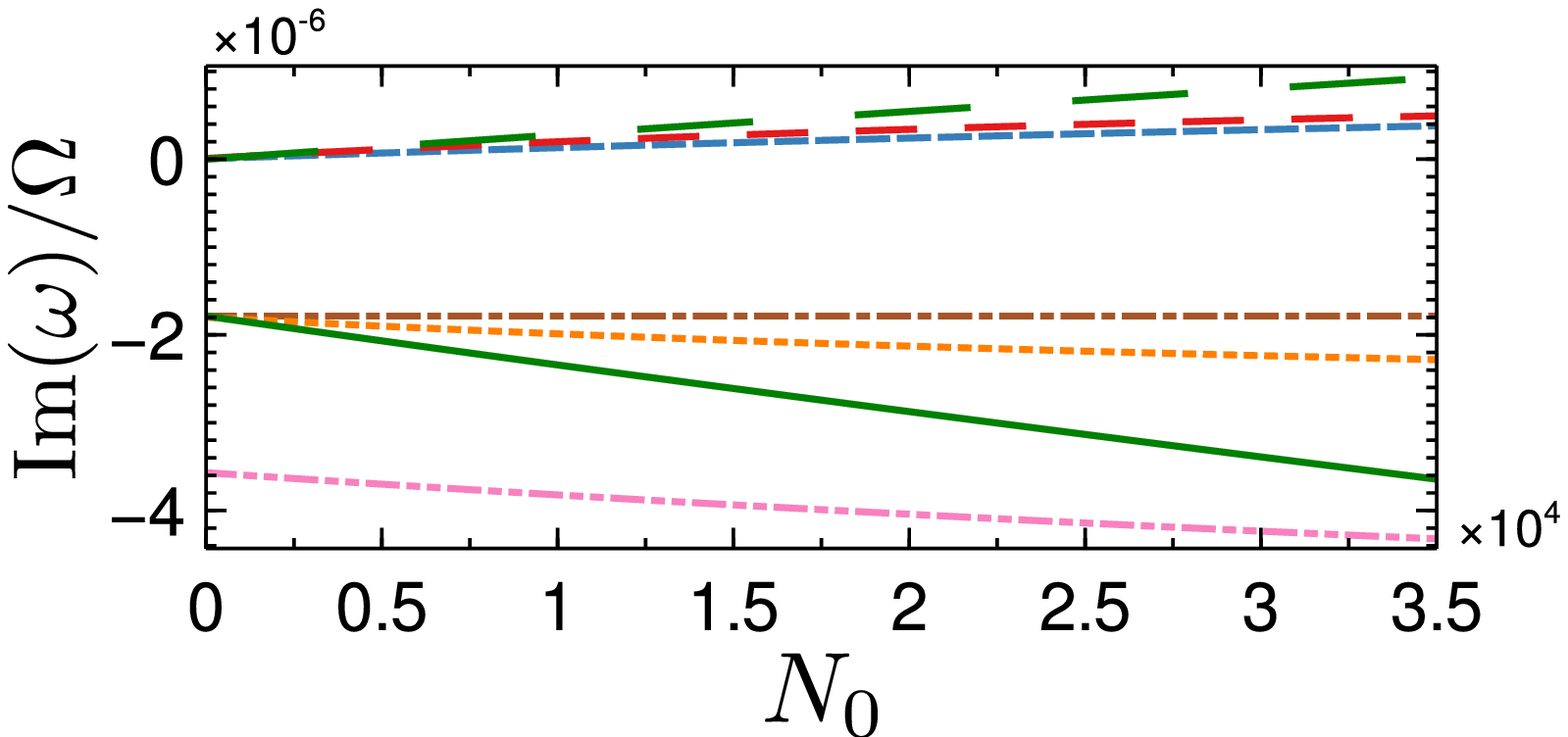}
	\caption{\label{Fig4}Damping rates of the eigenmodes of $S_\text{dipole}$ and $S_\text{widths}$. The green solid line corresponds to the damping of the dipole mode, whereas the green dashed line belongs to the temperature-difference oscillation. The remainder visualises the damping rates of the breathing mode (red, dashed), the quadrupole mode (blue, short dashed), the temperature-difference amplitude (brown, dashed-dotted), the temperature-difference breathing mode (orange, dotted) and one of the photon number (dashed-double-dotted).}
\end{figure}

In Fig.~\ref{Fig4} the damping rates corresponding to the oscillation frequencies from the main paper are plotted. As only a thermo-optic damping effect and no damping stemming from the matter is taken into account, these rates turn out to be quite small and also slightly positive. However, compared to the pulse duration, which is about \SI{500}{\nano\second}, the instability occurs on a much larger time scale of $10^{-6}/\Omega\sim10^4~ \si{\second}$. Nevertheless, this instability reflects the missing matter degree of freedom of the present mean-field theory.
 
\begin{thebibliography}{41}%
\makeatletter
\providecommand \@ifxundefined [1]{%
 \@ifx{#1\undefined}
}%
\providecommand \@ifnum [1]{%
 \ifnum #1\expandafter \@firstoftwo
 \else \expandafter \@secondoftwo
 \fi
}%
\providecommand \@ifx [1]{%
 \ifx #1\expandafter \@firstoftwo
 \else \expandafter \@secondoftwo
 \fi
}%
\providecommand \natexlab [1]{#1}%
\providecommand \enquote  [1]{``#1''}%
\providecommand \bibnamefont  [1]{#1}%
\providecommand \bibfnamefont [1]{#1}%
\providecommand \citenamefont [1]{#1}%
\providecommand \href@noop [0]{\@secondoftwo}%
\providecommand \href [0]{\begingroup \@sanitize@url \@href}%
\providecommand \@href[1]{\@@startlink{#1}\@@href}%
\providecommand \@@href[1]{\endgroup#1\@@endlink}%
\providecommand \@sanitize@url [0]{\catcode `\\12\catcode `\$12\catcode
  `\&12\catcode `\#12\catcode `\^12\catcode `\_12\catcode `\%12\relax}%
\providecommand \@@startlink[1]{}%
\providecommand \@@endlink[0]{}%
\providecommand \url  [0]{\begingroup\@sanitize@url \@url }%
\providecommand \@url [1]{\endgroup\@href {#1}{\urlprefix }}%
\providecommand \urlprefix  [0]{URL }%
\providecommand \Eprint [0]{\href }%
\providecommand \doibase [0]{http://dx.doi.org/}%
\providecommand \selectlanguage [0]{\@gobble}%
\providecommand \bibinfo  [0]{\@secondoftwo}%
\providecommand \bibfield  [0]{\@secondoftwo}%
\providecommand \translation [1]{[#1]}%
\providecommand \BibitemOpen [0]{}%
\providecommand \bibitemStop [0]{}%
\providecommand \bibitemNoStop [0]{.\EOS\space}%
\providecommand \EOS [0]{\spacefactor3000\relax}%
\providecommand \BibitemShut  [1]{\csname bibitem#1\endcsname}%
\let\auto@bib@innerbib\@empty
\bibitem [{\citenamefont {Carusotto}\ and\ \citenamefont
  {Ciuti}(2013)}]{RevModPhys.85.299}%
  \BibitemOpen
  \bibfield  {author} {\bibinfo {author} {\bibfnamefont {I.}~\bibnamefont
  {Carusotto}}\ and\ \bibinfo {author} {\bibfnamefont {C.}~\bibnamefont
  {Ciuti}},\ }\href {\doibase 10.1103/RevModPhys.85.299} {\bibfield  {journal}
  {\bibinfo  {journal} {Rev. Mod. Phys.}\ }\textbf {\bibinfo {volume} {85}},\
  \bibinfo {pages} {299} (\bibinfo {year} {2013})}\BibitemShut {NoStop}%
\bibitem [{\citenamefont {Lugiato}\ and\ \citenamefont
  {Lefever}(1987)}]{PhysRevLett.58.2209}%
  \BibitemOpen
  \bibfield  {author} {\bibinfo {author} {\bibfnamefont {L.~A.}\ \bibnamefont
  {Lugiato}}\ and\ \bibinfo {author} {\bibfnamefont {R.}~\bibnamefont
  {Lefever}},\ }\href {\doibase 10.1103/PhysRevLett.58.2209} {\bibfield
  {journal} {\bibinfo  {journal} {Phys. Rev. Lett.}\ }\textbf {\bibinfo
  {volume} {58}},\ \bibinfo {pages} {2209} (\bibinfo {year}
  {1987})}\BibitemShut {NoStop}%
\bibitem [{\citenamefont {Staliunas}(1993)}]{PhysRevA.48.1573}%
  \BibitemOpen
  \bibfield  {author} {\bibinfo {author} {\bibfnamefont {K.}~\bibnamefont
  {Staliunas}},\ }\href {\doibase 10.1103/PhysRevA.48.1573} {\bibfield
  {journal} {\bibinfo  {journal} {Phys. Rev. A}\ }\textbf {\bibinfo {volume}
  {48}},\ \bibinfo {pages} {1573} (\bibinfo {year} {1993})}\BibitemShut
  {NoStop}%
\bibitem [{\citenamefont {{G. A. Swartzlander}}\ and\ \citenamefont
  {Law}(1992)}]{PhysRevLett.69.2503}%
  \BibitemOpen
  \bibfield  {author} {\bibinfo {author} {\bibnamefont {{G.
  A. Swartzlander}}}\ and\ \bibinfo {author} {\bibfnamefont {C.~T.}\
  \bibnamefont {Law}},\ }\href@noop {} {\bibfield  {journal} {\bibinfo
  {journal} {Phys. Rev. Lett.}\ }\textbf {\bibinfo {volume} {69}},\ \bibinfo
  {pages} {2503} (\bibinfo {year} {1992})}\BibitemShut {NoStop}%
\bibitem [{\citenamefont {Deng}\ \emph {et~al.}(2002)\citenamefont {Deng},
  \citenamefont {Weihs}, \citenamefont {Santori}, \citenamefont {Bloch},\ and\
  \citenamefont {Yamamoto}}]{Science.298.5591}%
  \BibitemOpen
  \bibfield  {author} {\bibinfo {author} {\bibfnamefont {H.}~\bibnamefont
  {Deng}}, \bibinfo {author} {\bibfnamefont {G.}~\bibnamefont {Weihs}},
  \bibinfo {author} {\bibfnamefont {C.}~\bibnamefont {Santori}}, \bibinfo
  {author} {\bibfnamefont {J.}~\bibnamefont {Bloch}}, \ and\ \bibinfo {author}
  {\bibfnamefont {Y.}~\bibnamefont {Yamamoto}},\ }\href@noop {} {\bibfield
  {journal} {\bibinfo  {journal} {Science}\ }\textbf {\bibinfo {volume}
  {298}},\ \bibinfo {pages} {199} (\bibinfo {year} {2002})}\BibitemShut
  {NoStop}%
\bibitem [{\citenamefont {Klaers}\ \emph
  {et~al.}(2010{\natexlab{a}})\citenamefont {Klaers}, \citenamefont {Schmitt},
  \citenamefont {Vewinger},\ and\ \citenamefont {Weitz}}]{Klaers2010a}%
  \BibitemOpen
  \bibfield  {author} {\bibinfo {author} {\bibfnamefont {J.}~\bibnamefont
  {Klaers}}, \bibinfo {author} {\bibfnamefont {J.}~\bibnamefont {Schmitt}},
  \bibinfo {author} {\bibfnamefont {F.}~\bibnamefont {Vewinger}}, \ and\
  \bibinfo {author} {\bibfnamefont {M.}~\bibnamefont {Weitz}},\ }\href
  {\doibase 10.1038/nature09567} {\bibfield  {journal} {\bibinfo  {journal}
  {Nature}\ }\textbf {\bibinfo {volume} {468}},\ \bibinfo {pages} {545}
  (\bibinfo {year} {2010}{\natexlab{a}})}\BibitemShut {NoStop}%
\bibitem [{\citenamefont {Schmitt}\ \emph {et~al.}(2015)\citenamefont
  {Schmitt}, \citenamefont {Damm}, \citenamefont {Dung}, \citenamefont
  {Vewinger}, \citenamefont {Klaers},\ and\ \citenamefont
  {Weitz}}]{Schmitt2015}%
  \BibitemOpen
  \bibfield  {author} {\bibinfo {author} {\bibfnamefont {J.}~\bibnamefont
  {Schmitt}}, \bibinfo {author} {\bibfnamefont {T.}~\bibnamefont {Damm}},
  \bibinfo {author} {\bibfnamefont {D.}~\bibnamefont {Dung}}, \bibinfo {author}
  {\bibfnamefont {F.}~\bibnamefont {Vewinger}}, \bibinfo {author}
  {\bibfnamefont {J.}~\bibnamefont {Klaers}}, \ and\ \bibinfo {author}
  {\bibfnamefont {M.}~\bibnamefont {Weitz}},\ }\href {\doibase
  10.1103/PhysRevA.92.011602} {\bibfield  {journal} {\bibinfo  {journal} {Phys.
  Rev. A}\ }\textbf {\bibinfo {volume} {92}},\ \bibinfo {pages} {011602(R)}
  (\bibinfo {year} {2015})}\BibitemShut {NoStop}%
\bibitem [{\citenamefont {Klaers}\ \emph
  {et~al.}(2010{\natexlab{b}})\citenamefont {Klaers}, \citenamefont
  {Vewinger},\ and\ \citenamefont {Weitz}}]{Klaers2010}%
  \BibitemOpen
  \bibfield  {author} {\bibinfo {author} {\bibfnamefont {J.}~\bibnamefont
  {Klaers}}, \bibinfo {author} {\bibfnamefont {F.}~\bibnamefont {Vewinger}}, \
  and\ \bibinfo {author} {\bibfnamefont {M.}~\bibnamefont {Weitz}},\ }\href
  {\doibase 10.1038/nphys1680} {\bibfield  {journal} {\bibinfo  {journal} {Nat.
  Phys.}\ }\textbf {\bibinfo {volume} {6}},\ \bibinfo {pages} {512} (\bibinfo
  {year} {2010}{\natexlab{b}})}\BibitemShut {NoStop}%
\bibitem [{\citenamefont {Kennard}(1918)}]{Kennard1918}%
  \BibitemOpen
  \bibfield  {author} {\bibinfo {author} {\bibfnamefont {E.~H.}\ \bibnamefont
  {Kennard}},\ }\href {\doibase 10.1103/PhysRev.11.29} {\bibfield  {journal}
  {\bibinfo  {journal} {Phys. Rev.}\ }\textbf {\bibinfo {volume} {11}},\
  \bibinfo {pages} {29} (\bibinfo {year} {1918})}\BibitemShut {NoStop}%
\bibitem [{\citenamefont {Kennard}(1926)}]{Kennard1926}%
  \BibitemOpen
  \bibfield  {author} {\bibinfo {author} {\bibfnamefont {E.~H.}\ \bibnamefont
  {Kennard}},\ }\href {\doibase 10.1103/PhysRev.28.672} {\bibfield  {journal}
  {\bibinfo  {journal} {Phys. Rev.}\ }\textbf {\bibinfo {volume} {28}},\
  \bibinfo {pages} {672} (\bibinfo {year} {1926})}\BibitemShut {NoStop}%
\bibitem [{\citenamefont {Stepanov}(1957)}]{Stepanov1957}%
  \BibitemOpen
  \bibfield  {author} {\bibinfo {author} {\bibfnamefont {B.~I.}\ \bibnamefont
  {Stepanov}},\ }\href@noop {} {\bibfield  {journal} {\bibinfo  {journal}
  {Dokl. Akad. Nauk}\ }\textbf {\bibinfo {volume} {112}},\ \bibinfo {pages}
  {839} (\bibinfo {year} {1957})}\BibitemShut {NoStop}%
\bibitem [{\citenamefont {Kazachenko}\ and\ \citenamefont
  {Stepanov}(1957)}]{Kazachenko1957}%
  \BibitemOpen
  \bibfield  {author} {\bibinfo {author} {\bibfnamefont {L.~P.}\ \bibnamefont
  {Kazachenko}}\ and\ \bibinfo {author} {\bibfnamefont {B.~I.}\ \bibnamefont
  {Stepanov}},\ }\href@noop {} {\bibfield  {journal} {\bibinfo  {journal} {Opt.
  i Spektrosk.}\ }\textbf {\bibinfo {volume} {2}},\ \bibinfo {pages} {339}
  (\bibinfo {year} {1957})}\BibitemShut {NoStop}%
\bibitem [{\citenamefont {Klaers}\ \emph {et~al.}(2011)\citenamefont {Klaers},
  \citenamefont {Schmitt}, \citenamefont {Damm}, \citenamefont {Vewinger},\
  and\ \citenamefont {Weitz}}]{Klaers2011}%
  \BibitemOpen
  \bibfield  {author} {\bibinfo {author} {\bibfnamefont {J.}~\bibnamefont
  {Klaers}}, \bibinfo {author} {\bibfnamefont {J.}~\bibnamefont {Schmitt}},
  \bibinfo {author} {\bibfnamefont {T.}~\bibnamefont {Damm}}, \bibinfo {author}
  {\bibfnamefont {F.}~\bibnamefont {Vewinger}}, \ and\ \bibinfo {author}
  {\bibfnamefont {M.}~\bibnamefont {Weitz}},\ }\href {\doibase
  10.1007/s00340-011-4734-6} {\bibfield  {journal} {\bibinfo  {journal} {Appl.
  Phys. B}\ }\textbf {\bibinfo {volume} {105}},\ \bibinfo {pages} {17}
  (\bibinfo {year} {2011})}\BibitemShut {NoStop}%
\bibitem [{\citenamefont {Boyd}(2008)}]{Boyd}%
  \BibitemOpen
  \bibfield  {author} {\bibinfo {author} {\bibfnamefont {R.~W.}\ \bibnamefont
  {Boyd}},\ }\href
  {https://www.elsevier.com/books/nonlinear-optics/boyd/978-0-12-369470-6}
  {\emph {\bibinfo {title} {{Nonlinear Optics}}}},\ \bibinfo {edition} {3rd}\
  ed.\ (\bibinfo  {publisher} {Acadamic Press},\ \bibinfo {year}
  {2008})\BibitemShut {NoStop}%
\bibitem [{\citenamefont {Kirton}\ and\ \citenamefont
  {Keeling}(2013)}]{Kirton2013}%
  \BibitemOpen
  \bibfield  {author} {\bibinfo {author} {\bibfnamefont {P.}~\bibnamefont
  {Kirton}}\ and\ \bibinfo {author} {\bibfnamefont {J.}~\bibnamefont
  {Keeling}},\ }\href {\doibase 10.1103/PhysRevLett.111.100404} {\bibfield
  {journal} {\bibinfo  {journal} {Phys. Rev. Lett.}\ }\textbf {\bibinfo
  {volume} {111}},\ \bibinfo {pages} {100404} (\bibinfo {year}
  {2013})}\BibitemShut {NoStop}%
\bibitem [{\citenamefont {Radonji{\'{c}}}\ \emph {et~al.}(2018)\citenamefont
  {Radonji{\'{c}}}, \citenamefont {Kopylov}, \citenamefont {Bala{\v{z}}},\ and\
  \citenamefont {Pelster}}]{Radonjic2018}%
  \BibitemOpen
  \bibfield  {author} {\bibinfo {author} {\bibfnamefont {M.}~\bibnamefont
  {Radonji{\'{c}}}}, \bibinfo {author} {\bibfnamefont {W.}~\bibnamefont
  {Kopylov}}, \bibinfo {author} {\bibfnamefont {A.}~\bibnamefont
  {Bala{\v{z}}}}, \ and\ \bibinfo {author} {\bibfnamefont {A.}~\bibnamefont
  {Pelster}},\ }\href {\doibase 10.1088/1367-2630/aac2a6} {\bibfield  {journal}
  {\bibinfo  {journal} {New J. Phys.}\ }\textbf {\bibinfo {volume} {20}},\
  \bibinfo {pages} {055014} (\bibinfo {year} {2018})}\BibitemShut {NoStop}%
\bibitem [{\citenamefont {Marelic}\ \emph {et~al.}(2016)\citenamefont
  {Marelic}, \citenamefont {Walker},\ and\ \citenamefont
  {Nyman}}]{Marelic2016}%
  \BibitemOpen
  \bibfield  {author} {\bibinfo {author} {\bibfnamefont {J.}~\bibnamefont
  {Marelic}}, \bibinfo {author} {\bibfnamefont {B.~T.}\ \bibnamefont {Walker}},
  \ and\ \bibinfo {author} {\bibfnamefont {R.~A.}\ \bibnamefont {Nyman}},\
  }\href {\doibase 10.1103/PhysRevA.94.063812} {\bibfield  {journal} {\bibinfo
  {journal} {Phys. Rev. A}\ }\textbf {\bibinfo {volume} {94}},\ \bibinfo
  {pages} {063812} (\bibinfo {year} {2016})}\BibitemShut {NoStop}%
\bibitem [{\citenamefont {Stamper-Kurn}\ \emph {et~al.}(1998)\citenamefont
  {Stamper-Kurn}, \citenamefont {Miesner}, \citenamefont {Inouye},
  \citenamefont {Andrews},\ and\ \citenamefont
  {Ketterle}}]{PhysRevLett.81.500}%
  \BibitemOpen
  \bibfield  {author} {\bibinfo {author} {\bibfnamefont {D.~M.}\ \bibnamefont
  {Stamper-Kurn}}, \bibinfo {author} {\bibfnamefont {H.-J.}\ \bibnamefont
  {Miesner}}, \bibinfo {author} {\bibfnamefont {S.}~\bibnamefont {Inouye}},
  \bibinfo {author} {\bibfnamefont {M.~R.}\ \bibnamefont {Andrews}}, \ and\
  \bibinfo {author} {\bibfnamefont {W.}~\bibnamefont {Ketterle}},\ }\href
  {\doibase 10.1103/PhysRevLett.81.500} {\bibfield  {journal} {\bibinfo
  {journal} {Phys. Rev. Lett.}\ }\textbf {\bibinfo {volume} {81}},\ \bibinfo
  {pages} {500} (\bibinfo {year} {1998})}\BibitemShut {NoStop}%
\bibitem [{\citenamefont {Pollack}\ \emph {et~al.}(2010)\citenamefont
  {Pollack}, \citenamefont {Dries}, \citenamefont {Hulet}, \citenamefont
  {Magalh{\~{a}}es}, \citenamefont {Henn}, \citenamefont {Ramos}, \citenamefont
  {Caracanhas},\ and\ \citenamefont {Bagnato}}]{Pollack2010}%
  \BibitemOpen
  \bibfield  {author} {\bibinfo {author} {\bibfnamefont {S.~E.}\ \bibnamefont
  {Pollack}}, \bibinfo {author} {\bibfnamefont {D.}~\bibnamefont {Dries}},
  \bibinfo {author} {\bibfnamefont {R.~G.}\ \bibnamefont {Hulet}}, \bibinfo
  {author} {\bibfnamefont {K.~M.~F.}\ \bibnamefont {Magalh{\~{a}}es}}, \bibinfo
  {author} {\bibfnamefont {E.~A.~L.}\ \bibnamefont {Henn}}, \bibinfo {author}
  {\bibfnamefont {E.~R.~F.}\ \bibnamefont {Ramos}}, \bibinfo {author}
  {\bibfnamefont {M.~A.}\ \bibnamefont {Caracanhas}}, \ and\ \bibinfo {author}
  {\bibfnamefont {V.~S.}\ \bibnamefont {Bagnato}},\ }\href {\doibase
  10.1103/PhysRevA.81.053627} {\bibfield  {journal} {\bibinfo  {journal} {Phys.
  Rev. A}\ }\textbf {\bibinfo {volume} {81}},\ \bibinfo {pages} {053627}
  (\bibinfo {year} {2010})}\BibitemShut {NoStop}%
\bibitem [{\citenamefont {van~der Wurff}\ \emph {et~al.}(2014)\citenamefont
  {van~der Wurff}, \citenamefont {de~Leeuw}, \citenamefont {Duine},\ and\
  \citenamefont {Stoof}}]{PhysRevLett.113.135301}%
  \BibitemOpen
  \bibfield  {author} {\bibinfo {author} {\bibfnamefont {E.~C.~I.}\
  \bibnamefont {van~der Wurff}}, \bibinfo {author} {\bibfnamefont {A.-W.}\
  \bibnamefont {de~Leeuw}}, \bibinfo {author} {\bibfnamefont {R.~A.}\
  \bibnamefont {Duine}}, \ and\ \bibinfo {author} {\bibfnamefont {H.~T.~C.}\
  \bibnamefont {Stoof}},\ }\href {\doibase 10.1103/PhysRevLett.113.135301}
  {\bibfield  {journal} {\bibinfo  {journal} {Phys. Rev. Lett.}\ }\textbf
  {\bibinfo {volume} {113}},\ \bibinfo {pages} {135301} (\bibinfo {year}
  {2014})}\BibitemShut {NoStop}%
\bibitem [{\citenamefont {Alaeian}\ \emph {et~al.}(2017)\citenamefont
  {Alaeian}, \citenamefont {Schedensack}, \citenamefont {Bartels},
  \citenamefont {Peterseim},\ and\ \citenamefont {Weitz}}]{Alaeian2017}%
  \BibitemOpen
  \bibfield  {author} {\bibinfo {author} {\bibfnamefont {H.}~\bibnamefont
  {Alaeian}}, \bibinfo {author} {\bibfnamefont {M.}~\bibnamefont
  {Schedensack}}, \bibinfo {author} {\bibfnamefont {C.}~\bibnamefont
  {Bartels}}, \bibinfo {author} {\bibfnamefont {D.}~\bibnamefont {Peterseim}},
  \ and\ \bibinfo {author} {\bibfnamefont {M.}~\bibnamefont {Weitz}},\ }\href
  {\doibase 10.1088/1367-2630/aa964c} {\bibfield  {journal} {\bibinfo
  {journal} {New J. Phys.}\ }\textbf {\bibinfo {volume} {19}},\ \bibinfo
  {pages} {115009} (\bibinfo {year} {2017})}\BibitemShut {NoStop}%
\bibitem [{\citenamefont {Dung}\ \emph {et~al.}(2017)\citenamefont {Dung},
  \citenamefont {Kurtscheid}, \citenamefont {Damm}, \citenamefont {Schmitt},
  \citenamefont {Vewinger}, \citenamefont {Weitz},\ and\ \citenamefont
  {Klaers}}]{Dung2017}%
  \BibitemOpen
  \bibfield  {author} {\bibinfo {author} {\bibfnamefont {D.}~\bibnamefont
  {Dung}}, \bibinfo {author} {\bibfnamefont {C.}~\bibnamefont {Kurtscheid}},
  \bibinfo {author} {\bibfnamefont {T.}~\bibnamefont {Damm}}, \bibinfo {author}
  {\bibfnamefont {J.}~\bibnamefont {Schmitt}}, \bibinfo {author} {\bibfnamefont
  {F.}~\bibnamefont {Vewinger}}, \bibinfo {author} {\bibfnamefont
  {M.}~\bibnamefont {Weitz}}, \ and\ \bibinfo {author} {\bibfnamefont
  {J.}~\bibnamefont {Klaers}},\ }\href {\doibase 10.1038/nphoton.2017.139}
  {\bibfield  {journal} {\bibinfo  {journal} {Nat. Photonics}\ }\textbf
  {\bibinfo {volume} {11}},\ \bibinfo {pages} {565} (\bibinfo {year}
  {2017})}\BibitemShut {NoStop}%
\bibitem [{\citenamefont {Lax}\ \emph {et~al.}(1975)\citenamefont {Lax},
  \citenamefont {Louisell},\ and\ \citenamefont {McKnight}}]{PhysRevA.11.1365}%
  \BibitemOpen
  \bibfield  {author} {\bibinfo {author} {\bibfnamefont {M.}~\bibnamefont
  {Lax}}, \bibinfo {author} {\bibfnamefont {W.~H.}\ \bibnamefont {Louisell}}, \
  and\ \bibinfo {author} {\bibfnamefont {W.~B.}\ \bibnamefont {McKnight}},\
  }\href {\doibase 10.1103/PhysRevA.11.1365} {\bibfield  {journal} {\bibinfo
  {journal} {Phys. Rev. A}\ }\textbf {\bibinfo {volume} {11}},\ \bibinfo
  {pages} {1365} (\bibinfo {year} {1975})}\BibitemShut {NoStop}%
\bibitem [{\citenamefont {Nyman}\ and\ \citenamefont
  {Walker}(2018)}]{Nyman2018}%
  \BibitemOpen
  \bibfield  {author} {\bibinfo {author} {\bibfnamefont {R.~A.}\ \bibnamefont
  {Nyman}}\ and\ \bibinfo {author} {\bibfnamefont {B.~T.}\ \bibnamefont
  {Walker}},\ }\href {\doibase 10.1080/09500340.2017.1404655} {\bibfield
  {journal} {\bibinfo  {journal} {J. Mod. Opt.}\ }\textbf {\bibinfo {volume}
  {65}},\ \bibinfo {pages} {754} (\bibinfo {year} {2018})}\BibitemShut
  {NoStop}%
\bibitem [{\citenamefont {{Calvanese Strinati}}\ and\ \citenamefont
  {Conti}(2014)}]{PhysRevA.90.043853}%
  \BibitemOpen
  \bibfield  {author} {\bibinfo {author} {\bibfnamefont {M.}~\bibnamefont
  {{Calvanese Strinati}}}\ and\ \bibinfo {author} {\bibfnamefont
  {C.}~\bibnamefont {Conti}},\ }\href {\doibase 10.1103/PhysRevA.90.043853}
  {\bibfield  {journal} {\bibinfo  {journal} {Phys. Rev. A}\ }\textbf {\bibinfo
  {volume} {90}},\ \bibinfo {pages} {43853} (\bibinfo {year}
  {2014})}\BibitemShut {NoStop}%
\bibitem [{\citenamefont {Stein}(2018)}]{Stein2018}%
  \BibitemOpen
  \bibfield  {author} {\bibinfo {author} {\bibfnamefont {E.}~\bibnamefont
  {Stein}},\ }\emph {\bibinfo {title} {{Open-Dissipative Mean-Field Theory for
  Photon Bose-Einstein Condensates}}},\ \href
  {http://users.physik.fu-berlin.de/~pelster/Theses/stein.pdf} {\bibinfo {type}
  {Diploma thesis}},\ \bibinfo  {school} {Technische Universit{\"{a}}t
  Kaiserslautern} (\bibinfo {year} {2018})\BibitemShut {NoStop}%
\bibitem [{\citenamefont {Wouters}\ and\ \citenamefont
  {Carusotto}(2007)}]{PhysRevLett.99.140402}%
  \BibitemOpen
  \bibfield  {author} {\bibinfo {author} {\bibfnamefont {M.}~\bibnamefont
  {Wouters}}\ and\ \bibinfo {author} {\bibfnamefont {I.}~\bibnamefont
  {Carusotto}},\ }\href {\doibase 10.1103/PhysRevLett.99.140402} {\bibfield
  {journal} {\bibinfo  {journal} {Phys. Rev. Lett.}\ }\textbf {\bibinfo
  {volume} {99}},\ \bibinfo {pages} {140402} (\bibinfo {year}
  {2007})}\BibitemShut {NoStop}%
\bibitem [{Sup()}]{Suppl}%
  \BibitemOpen
  \href@noop {} {}\bibinfo {howpublished} {See supplemental
  material}\BibitemShut {NoStop}%
\bibitem [{\citenamefont {Landau}\ and\ \citenamefont
  {Lif{\v{s}}ic}(2009)}]{Landau2009}%
  \BibitemOpen
  \bibfield  {author} {\bibinfo {author} {\bibfnamefont {L.~D.}\ \bibnamefont
  {Landau}}\ and\ \bibinfo {author} {\bibfnamefont {E.~M.}\ \bibnamefont
  {Lif{\v{s}}ic}},\ }\href@noop {} {\emph {\bibinfo {title} {{Course of
  Theoretical Physics}}}},\ \bibinfo {edition} {2nd}\ ed.\ (\bibinfo
  {publisher} {Elsevier},\ \bibinfo {address} {Amsterdam},\ \bibinfo {year}
  {2009})\BibitemShut {NoStop}%
\bibitem [{\citenamefont {P{\'{e}}rez-Garc{\'{i}}a}\ \emph
  {et~al.}(1996)\citenamefont {P{\'{e}}rez-Garc{\'{i}}a}, \citenamefont
  {Michinel}, \citenamefont {Cirac}, \citenamefont {Lewenstein},\ and\
  \citenamefont {Zoller}}]{PhysRevLett.77.5320}%
  \BibitemOpen
  \bibfield  {author} {\bibinfo {author} {\bibfnamefont {V.~M.}\ \bibnamefont
  {P{\'{e}}rez-Garc{\'{i}}a}}, \bibinfo {author} {\bibfnamefont
  {H.}~\bibnamefont {Michinel}}, \bibinfo {author} {\bibfnamefont {J.~I.}\
  \bibnamefont {Cirac}}, \bibinfo {author} {\bibfnamefont {M.}~\bibnamefont
  {Lewenstein}}, \ and\ \bibinfo {author} {\bibfnamefont {P.}~\bibnamefont
  {Zoller}},\ }\href {\doibase 10.1103/PhysRevLett.77.5320} {\bibfield
  {journal} {\bibinfo  {journal} {Phys. Rev. Lett.}\ }\textbf {\bibinfo
  {volume} {77}},\ \bibinfo {pages} {5320} (\bibinfo {year}
  {1996})}\BibitemShut {NoStop}%
\bibitem [{\citenamefont {P{\'{e}}rez-Garc{\'{i}}a}\ \emph
  {et~al.}(1997)\citenamefont {P{\'{e}}rez-Garc{\'{i}}a}, \citenamefont
  {Michinel}, \citenamefont {Cirac}, \citenamefont {Lewenstein},\ and\
  \citenamefont {Zoller}}]{PhysRevA.56.1424}%
  \BibitemOpen
  \bibfield  {author} {\bibinfo {author} {\bibfnamefont {V.~M.}\ \bibnamefont
  {P{\'{e}}rez-Garc{\'{i}}a}}, \bibinfo {author} {\bibfnamefont
  {H.}~\bibnamefont {Michinel}}, \bibinfo {author} {\bibfnamefont {J.~I.}\
  \bibnamefont {Cirac}}, \bibinfo {author} {\bibfnamefont {M.}~\bibnamefont
  {Lewenstein}}, \ and\ \bibinfo {author} {\bibfnamefont {P.}~\bibnamefont
  {Zoller}},\ }\href {\doibase 10.1103/PhysRevA.56.1424} {\bibfield  {journal}
  {\bibinfo  {journal} {Phys. Rev. A}\ }\textbf {\bibinfo {volume} {56}},\
  \bibinfo {pages} {1424} (\bibinfo {year} {1997})}\BibitemShut {NoStop}%
\bibitem [{\citenamefont {Fetter}\ and\ \citenamefont
  {Rokhsar}(1998)}]{PhysRevA.57.1191}%
  \BibitemOpen
  \bibfield  {author} {\bibinfo {author} {\bibfnamefont {A.~L.}\ \bibnamefont
  {Fetter}}\ and\ \bibinfo {author} {\bibfnamefont {D.}~\bibnamefont
  {Rokhsar}},\ }\href {\doibase 10.1103/PhysRevA.57.1191} {\bibfield  {journal}
  {\bibinfo  {journal} {Phys. Rev. A}\ }\textbf {\bibinfo {volume} {57}},\
  \bibinfo {pages} {1191} (\bibinfo {year} {1998})}\BibitemShut {NoStop}%
\bibitem [{\citenamefont {Busch}\ \emph {et~al.}(1997)\citenamefont {Busch},
  \citenamefont {Cirac}, \citenamefont {P{\'{e}}rez-Garc{\'{i}}a},\ and\
  \citenamefont {Zoller}}]{PhysRevA.56.2978}%
  \BibitemOpen
  \bibfield  {author} {\bibinfo {author} {\bibfnamefont {T.}~\bibnamefont
  {Busch}}, \bibinfo {author} {\bibfnamefont {J.~I.}\ \bibnamefont {Cirac}},
  \bibinfo {author} {\bibfnamefont {V.~M.}\ \bibnamefont
  {P{\'{e}}rez-Garc{\'{i}}a}}, \ and\ \bibinfo {author} {\bibfnamefont
  {P.}~\bibnamefont {Zoller}},\ }\href {\doibase 10.1103/PhysRevA.56.2978}
  {\bibfield  {journal} {\bibinfo  {journal} {Phys. Rev. A}\ }\textbf {\bibinfo
  {volume} {56}},\ \bibinfo {pages} {2978} (\bibinfo {year}
  {1997})}\BibitemShut {NoStop}%
\bibitem [{\citenamefont {Mann}\ \emph {et~al.}(2018)\citenamefont {Mann},
  \citenamefont {Bakhtiari}, \citenamefont {Pelster},\ and\ \citenamefont
  {Thorwart}}]{PhysRevLett.120.063605}%
  \BibitemOpen
  \bibfield  {author} {\bibinfo {author} {\bibfnamefont {N.}~\bibnamefont
  {Mann}}, \bibinfo {author} {\bibfnamefont {M.~R.}\ \bibnamefont {Bakhtiari}},
  \bibinfo {author} {\bibfnamefont {A.}~\bibnamefont {Pelster}}, \ and\
  \bibinfo {author} {\bibfnamefont {M.}~\bibnamefont {Thorwart}},\ }\href
  {\doibase 10.1103/PhysRevLett.120.063605} {\bibfield  {journal} {\bibinfo
  {journal} {Phys. Rev. Lett.}\ }\textbf {\bibinfo {volume} {120}},\ \bibinfo
  {pages} {063605} (\bibinfo {year} {2018})}\BibitemShut {NoStop}%
\bibitem [{\citenamefont {Muruganandam}\ and\ \citenamefont
  {Adhikari}(2009)}]{Muruganandam2009}%
  \BibitemOpen
  \bibfield  {author} {\bibinfo {author} {\bibfnamefont {P.}~\bibnamefont
  {Muruganandam}}\ and\ \bibinfo {author} {\bibfnamefont {S.}~\bibnamefont
  {Adhikari}},\ }\href {\doibase 10.1016/j.cpc.2009.04.015} {\bibfield
  {journal} {\bibinfo  {journal} {Comput. Phys. Commun.}\ }\textbf {\bibinfo
  {volume} {180}},\ \bibinfo {pages} {1888} (\bibinfo {year}
  {2009})}\BibitemShut {NoStop}%
\bibitem [{\citenamefont {Vudragovi{\'{c}}}\ \emph {et~al.}(2012)\citenamefont
  {Vudragovi{\'{c}}}, \citenamefont {Vidanovi{\'{c}}}, \citenamefont
  {Bala{\v{z}}}, \citenamefont {Muruganandam},\ and\ \citenamefont
  {Adhikari}}]{Vudragovic2012}%
  \BibitemOpen
  \bibfield  {author} {\bibinfo {author} {\bibfnamefont {D.}~\bibnamefont
  {Vudragovi{\'{c}}}}, \bibinfo {author} {\bibfnamefont {I.}~\bibnamefont
  {Vidanovi{\'{c}}}}, \bibinfo {author} {\bibfnamefont {A.}~\bibnamefont
  {Bala{\v{z}}}}, \bibinfo {author} {\bibfnamefont {P.}~\bibnamefont
  {Muruganandam}}, \ and\ \bibinfo {author} {\bibfnamefont {S.~K.}\
  \bibnamefont {Adhikari}},\ }\href {\doibase 10.1016/j.cpc.2012.03.022}
  {\bibfield  {journal} {\bibinfo  {journal} {Comput. Phys. Commun.}\ }\textbf
  {\bibinfo {volume} {183}},\ \bibinfo {pages} {2021} (\bibinfo {year}
  {2012})}\BibitemShut {NoStop}%
\bibitem [{\citenamefont {Young-S.}\ \emph {et~al.}(2016)\citenamefont
  {Young-S.}, \citenamefont {Vudragovi{\'{c}}}, \citenamefont {Muruganandam},
  \citenamefont {Adhikari},\ and\ \citenamefont {Bala{\v{z}}}}]{Young-S.2016}%
  \BibitemOpen
  \bibfield  {author} {\bibinfo {author} {\bibfnamefont {L.~E.}\ \bibnamefont
  {Young-S.}}, \bibinfo {author} {\bibfnamefont {D.}~\bibnamefont
  {Vudragovi{\'{c}}}}, \bibinfo {author} {\bibfnamefont {P.}~\bibnamefont
  {Muruganandam}}, \bibinfo {author} {\bibfnamefont {S.~K.}\ \bibnamefont
  {Adhikari}}, \ and\ \bibinfo {author} {\bibfnamefont {A.}~\bibnamefont
  {Bala{\v{z}}}},\ }\href {\doibase 10.1016/j.cpc.2016.03.015} {\bibfield
  {journal} {\bibinfo  {journal} {Comput. Phys. Commun.}\ }\textbf {\bibinfo
  {volume} {204}},\ \bibinfo {pages} {209} (\bibinfo {year}
  {2016})}\BibitemShut {NoStop}%
\bibitem [{\citenamefont {Ghosh}\ and\ \citenamefont
  {Sinha}(2002)}]{Ghosh2002}%
  \BibitemOpen
  \bibfield  {author} {\bibinfo {author} {\bibfnamefont {T.}~\bibnamefont
  {Ghosh}}\ and\ \bibinfo {author} {\bibfnamefont {S.}~\bibnamefont {Sinha}},\
  }\href {\doibase 10.1140/epjd/e20020086} {\bibfield  {journal} {\bibinfo
  {journal} {Eur. Phys. J. D - At. Mol. Opt. Phys.}\ }\textbf {\bibinfo
  {volume} {19}},\ \bibinfo {pages} {371} (\bibinfo {year} {2002})}\BibitemShut
  {NoStop}%
\bibitem [{\citenamefont {Nyman}(2017)}]{nyman_robert_andrew_2017_569817}%
  \BibitemOpen
  \bibfield  {author} {\bibinfo {author} {\bibfnamefont {R.~A.}\ \bibnamefont
  {Nyman}},\ }\href {\doibase 10.5281/zenodo.569817} {}\bibinfo {howpublished}
  {Absorption and Fluorescence spectra of Rhodamine 6G} (\bibinfo {year}
  {2017}),\ \Eprint {http://arxiv.org/abs/https://zenodo.org/record/569817}
  {https://zenodo.org/record/569817} \BibitemShut {NoStop}%
\bibitem [{eth()}]{ethyleneglycol}%
  \BibitemOpen
  \href@noop {} {}\bibinfo {howpublished} {Ethylene Glycol, MEGlobal Product
  Guide},\ \Eprint
  {http://arxiv.org/abs/{www.meglobal.biz/media/product\_guides/MEGlobal\_MEG.pdf}}
  {{www.meglobal.biz/media/product\_guides/MEGlobal\_MEG.pdf}} \BibitemShut
  {NoStop}%
\bibitem [{\citenamefont {Nyman}()}]{PrivNyman}%
  \BibitemOpen
  \bibfield  {author} {\bibinfo {author} {\bibfnamefont {R.~A.}\ \bibnamefont
  {Nyman}},\ }\href@noop {} {}\bibinfo {howpublished} {Private
  Communication}\BibitemShut {NoStop}%
\end{thebibliography}

\begin{thebibliography}{2}%
\makeatletter
\providecommand \@ifxundefined [1]{%
 \@ifx{#1\undefined}
}%
\providecommand \@ifnum [1]{%
 \ifnum #1\expandafter \@firstoftwo
 \else \expandafter \@secondoftwo
 \fi
}%
\providecommand \@ifx [1]{%
 \ifx #1\expandafter \@firstoftwo
 \else \expandafter \@secondoftwo
 \fi
}%
\providecommand \natexlab [1]{#1}%
\providecommand \enquote  [1]{``#1''}%
\providecommand \bibnamefont  [1]{#1}%
\providecommand \bibfnamefont [1]{#1}%
\providecommand \citenamefont [1]{#1}%
\providecommand \href@noop [0]{\@secondoftwo}%
\providecommand \href [0]{\begingroup \@sanitize@url \@href}%
\providecommand \@href[1]{\@@startlink{#1}\@@href}%
\providecommand \@@href[1]{\endgroup#1\@@endlink}%
\providecommand \@sanitize@url [0]{\catcode `\\12\catcode `\$12\catcode
  `\&12\catcode `\#12\catcode `\^12\catcode `\_12\catcode `\%12\relax}%
\providecommand \@@startlink[1]{}%
\providecommand \@@endlink[0]{}%
\providecommand \url  [0]{\begingroup\@sanitize@url \@url }%
\providecommand \@url [1]{\endgroup\@href {#1}{\urlprefix }}%
\providecommand \urlprefix  [0]{URL }%
\providecommand \Eprint [0]{\href }%
\providecommand \doibase [0]{http://dx.doi.org/}%
\providecommand \selectlanguage [0]{\@gobble}%
\providecommand \bibinfo  [0]{\@secondoftwo}%
\providecommand \bibfield  [0]{\@secondoftwo}%
\providecommand \translation [1]{[#1]}%
\providecommand \BibitemOpen [0]{}%
\providecommand \bibitemStop [0]{}%
\providecommand \bibitemNoStop [0]{.\EOS\space}%
\providecommand \EOS [0]{\spacefactor3000\relax}%
\providecommand \BibitemShut  [1]{\csname bibitem#1\endcsname}%
\let\auto@bib@innerbib\@empty
\bibitem [{\citenamefont {Murray}(2002)}]{Murray2002a}%
  \BibitemOpen
  \bibfield  {author} {\bibinfo {author} {\bibfnamefont {J.~D.}\ \bibnamefont
  {Murray}},\ }\href {https://kplus.ub.uni-kl.de/Record/KLU01-000837584} {\emph
  {\bibinfo {title} {{Mathematical biology}}}},\ \bibinfo {edition} {corr. 2.
  p}\ ed.,\ Interdisciplinary applied mathematics; 18\ (\bibinfo  {publisher}
  {Springer, Berlin},\ \bibinfo {year} {2002})\BibitemShut {NoStop}%
\bibitem [{\citenamefont {Mann}\ \emph {et~al.}(2018)\citenamefont {Mann},
  \citenamefont {Bakhtiari}, \citenamefont {Pelster},\ and\ \citenamefont
  {Thorwart}}]{PhysRevLett.120.063605}%
  \BibitemOpen
  \bibfield  {author} {\bibinfo {author} {\bibfnamefont {N.}~\bibnamefont
  {Mann}}, \bibinfo {author} {\bibfnamefont {M.~R.}\ \bibnamefont {Bakhtiari}},
  \bibinfo {author} {\bibfnamefont {A.}~\bibnamefont {Pelster}}, \ and\
  \bibinfo {author} {\bibfnamefont {M.}~\bibnamefont {Thorwart}},\ }\href
  {\doibase 10.1103/PhysRevLett.120.063605} {\bibfield  {journal} {\bibinfo
  {journal} {Phys. Rev. Lett.}\ }\textbf {\bibinfo {volume} {120}},\ \bibinfo
  {pages} {063605} (\bibinfo {year} {2018})}\BibitemShut {NoStop}%
\end{thebibliography}
\end{document}